\documentclass[superscriptaddress,onecolumn,aps,pra]{revtex4-2}
\usepackage[utf8]{inputenc}

\usepackage{amsmath,amssymb,amsfonts}
\usepackage{bm}
\usepackage{graphicx}
\usepackage{fancyhdr}
\usepackage{xcolor}
\usepackage{hyperref}
\usepackage{enumitem}
\usepackage[normalem]{ulem}
\usepackage[caption=false]{subfig}
\usepackage{graphicx}
\usepackage{float} % Ensures proper float handling

\usepackage{fancyhdr}
\newcommand{\e}{\bm{\hat{e}}}

\newcommand{\nagn}{\bm{n}}

\newcommand{\heff}{\bm{f}}
\newcommand{\hext}{h}
\newcommand{\bhext}{\bm{h}}

\newcommand{\dm}{\lambda}

\newcommand{\lw}{\ell_w}

\newcommand{\Eeff}{F}

\begin{document}

\title{Traveling antiferromagnetic domain walls in a magnetic field}

\author{G. Theodorou}
\affiliation{Department of Mathematics and Applied Mathematics, University of Crete, 70013 Heraklion, Crete, Greece}
\affiliation{Institute of Applied and Computational Mathematics, FORTH, Heraklion, Crete, Greece}
\author{S. Komineas}
\affiliation{Department of Mathematics and Applied Mathematics, University of Crete, 70013 Heraklion, Crete, Greece}
\affiliation{Institute of Applied and Computational Mathematics, FORTH, Heraklion, Crete, Greece}
\date{\today}

\begin{abstract}
We consider an antiferromagnet in one space dimension with easy-axis anisotropy in a perpendicular magnetic field.
We study propagating domain wall solutions that can have a velocity up to a maximum $v_c$.
The width of the domain wall is a non-monotonic function of the velocity and it diverges to infinity at $v_c$.
Both features are in contrast to the case of the Lorentz invariant model in the absence of the field.
We further study the modification of the wall profile when a Dzyaloshinskii-Moriya interaction is added.
Finally, we present a propagating spiral expected to form when the system is forced at a velocity higher than the maximum velocity for domain walls and we give numerical results for the effect of the Dzyaloshinskii-Moriya interaction.
\end{abstract}

\maketitle

\section{Introduction}
\label{sec:introduction}

Antiferromagnetic domain walls have attracted attention for many decades \cite{1990_PR_KosevichIvanovKovalev,1991_AdvPhys_MikeskaSteiner} while interest has grown in recent years as antiferromagnetic order can be observed in a more efficient way and domain walls can now be directly imaged  \cite{2020_npjQM_CheongFiebig,2021_NatPhys_HedrichShekaMakarov,2022_acsnano_GrigorevKlaeuiDemsar}.
In addition, experimental techniques allow for optical \cite{2023_advMater_MeerKlaeui} and electrical \cite{2018_NatNano_WadleyJungwidth} manipulation of domain walls (DW) in antiferromagnets (AFMs), while current-induced switching can be achieved by domain wall motion \cite{2019_PRL_BaldratiKlaeui}.

The most attractive features of antiferromagnets are related to their dynamics.
This is described by extensions of the nonlinear $\sigma$ model for the N\'eel vector \cite{1990_PR_KosevichIvanovKovalev,1991_AdvPhys_MikeskaSteiner} and thus is radically different from the dynamics of magnetization in ferromagnets.
An advantage of AFMs is fast dynamics, such as THz frequencies and fast domain wall motion \cite{2011_PRL_HalsBrataas,2016_PRL_ShiinoLee,2016_PRL_GomonayJungwirthSinova}.

The features and properties of antiferromagnetic domain walls have been theoretically studied in various cases.
It has been proposed that dislocations are sources of domain walls. \cite{1977_SJLTP_KovalevKosevich, 1977_JETPLett_Dzyaloshinskii,2000_LTP_DudkoKovalev}.
Internal dynamical modes of the domain walls have been studied in an easy-plane AFM in a strong in-plane magnetic field \cite{1997_PRB_IvanovKolezhuk} and in biaxial antiferromagnets \cite{1999_LTP_Kovalev,2018_LTP_Kovalev}, and the dynamic structure factor was calculated for a moving domain wall in an easy-axis antiferromagnet under an external field \cite{1991_EPL_IvanovKolezhuk}.

In the case where an external magnetic field is applied, the sigma model is substantially modified, and the Lorentz invariance is broken.
Despite this complication, static and propagating domain wall solutions are known to exist also in this case.
We focus on traveling domain wall solutions in an external field applied parallel to the easy axis of the antiferromagnet and found in \cite{1979_SJLTP_BaryakhtarIvanov,1979_SSC_BaryakhtarIvanov}.
We study its features and the unusual behavior of the domain wall width as a function of its velocity.
This dependence can be non-monotonic, while there is a maximum speed.
The wall width diverges for large speeds in stark contrast to the usual behavior of collapse for large speeds in the case of the Lorentz invariant model (when no external field is present).
We extend the results to the case where a Dzyaloshinskii-Moriya (DM) interaction is present.
We further investigate the fate of the system when the wall velocity exceeds the critical value and the domain wall disappears.
We find that a propagating spiral state is then the appropriate solution of the model.

The paper is organized as follows.
Sec.~\ref{sec:travelingWalls} gives the formalism and describes the propagating domain wall under an external field without and with a DM interaction, in two subsections.
Sec.~\ref{sec:spiral} describes the propagating spiral.
Sec.~\ref{sec:conclusions} contains our concluding remarks.

\section{Traveling domain walls}
\label{sec:travelingWalls}

We consider an antiferromagnet with easy-axis anisotropy and DM interaction in an external field $\bhext$.
In the continuum limit, statics and dynamics can be described by a nonlinear $\sigma$-model for the N\'eel vector $\nagn$ \cite{1990_PR_KosevichIvanovKovalev,1991_AdvPhys_MikeskaSteiner},
\begin{equation} \label{eq:sigmaModel_1D}
    \nagn\times (\ddot{\nagn} + 2\bhext\times\dot{\nagn} - \heff) = 0,\qquad \heff = \nagn'' - 2\dm\,\e_2\times\nagn' + n_3\e_3 - (\nagn\cdot\bhext)\bhext
\end{equation}
where the dot denotes differentiation in time, the prime denotes differentiation in the space variable $x$, and $\dm$ is the scaled DM parameter.
Throughout this work, we choose an external magnetic field parallel to the easy axis,
\[
\bhext = (0,0,\hext),
\]
Then, the model reduces to
\begin{equation} \label{eq:sigmaModel_1D_h3}
    \nagn\times (\ddot{\nagn} + 2\hext\,\e_3\times\dot{\nagn} - \heff ) = 0,\qquad \heff = \nagn'' - 2\dm\,\e_2\times\nagn' + (1-\hext^2) n_3\e_3.
\end{equation}
We will study traveling waves, that is, solutions of the form
\[
\nagn(x,t) = \nagn(\xi),\qquad \xi=x-vt.
\]
This form is used in Eq.~\eqref{eq:sigmaModel_1D_h3} which reduces to
\begin{equation} \label{eq:sigmaModel_1D_traveling}
    \nagn\times \left[ (1-v^2) \nagn'' + 2\hext v\,\e_3\times\nagn' - 2\dm\,\e_2\times\nagn' + (1-\hext^2) n_3\e_3\right] = 0.
\end{equation}

\subsection{Effect of the external field}

We start the study assuming $\dm=0$ (no DM interaction).
Equation~\eqref{eq:sigmaModel_1D} has a static domain wall solution.
In the absence of an external field, $\hext=0$, this can be made dynamical by a Lorentz transformation.
Lorentz invariance is lost once a field is applied, $\hext\neq 0$.
We will re-derive the dynamical domain wall solution, in the presence of an external field, that has been obtained in Ref.~\cite{1979_SJLTP_BaryakhtarIvanov}, and we will study its features.

All subsequent calculations will be based on the spherical parametrization for the N\'eel vector,
\begin{equation} \label{eq:sphericalParametrization}
n_1 = \sin\Theta \cos\Phi,\quad n_2 = \sin\Theta \sin\Phi,\quad n_3 = \cos\Theta.
\end{equation}
Equation~\eqref{eq:sigmaModel_1D_traveling} gives the system
\begin{subequations} \label{eq:ThetaPhi_traveling}
\begin{align}
    & (1-v^2)\Theta''-\left[2 \hext v \Phi' +(1-v^2)\Phi'^2+1-h^2\right] \sin \Theta\cos\Theta = 0 \label{eq:ThetaPhi_traveling_Theta}\\
    & \sin\Theta (1-v^2)\Phi'' +2 \cos \Theta \Theta' \left[ (1-v^2)\Phi'+\hext v\right] = 0. \label{eq:ThetaPhi_traveling_Phi}
\end{align}
\end{subequations}

Equation~\eqref{eq:ThetaPhi_traveling_Phi} gives
\begin{equation} \label{eq:Phi_integrated}
 \left[ \sin^2\Theta \left[(1-v^2) \Phi' + \hext v \right] \right]' = 0
 \Rightarrow (1-v^2) \Phi' + \hext v = 0.
\end{equation} 
The integration constant was set to zero because we are looking for domain walls and thus have $\Theta=0,\pi$ at the two ends of the system.
The solution is
\begin{equation} \label{eq:PhiSol}
  \Phi=k\left(\xi-x_1\right), \qquad k = -\frac{hv}{1-v^2}
\end{equation}
where $x_1$ is an arbitrary constant, such that $\Phi(x=x_1,t=0)=0$.

\begin{figure}[t]
    \centering
\subfloat[]
{\includegraphics[width=6cm]{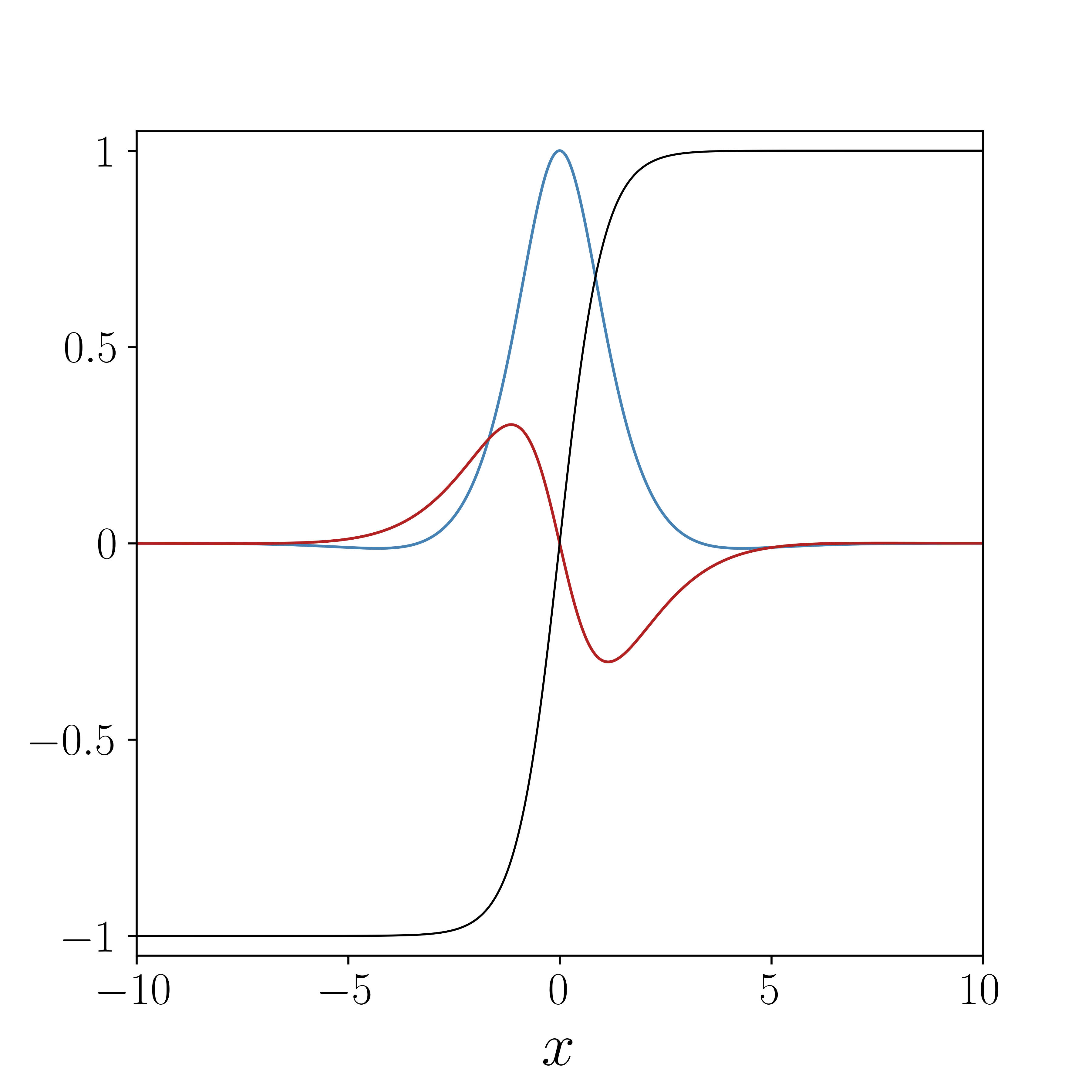}}\qquad
\subfloat[]
{\includegraphics[width=6cm]{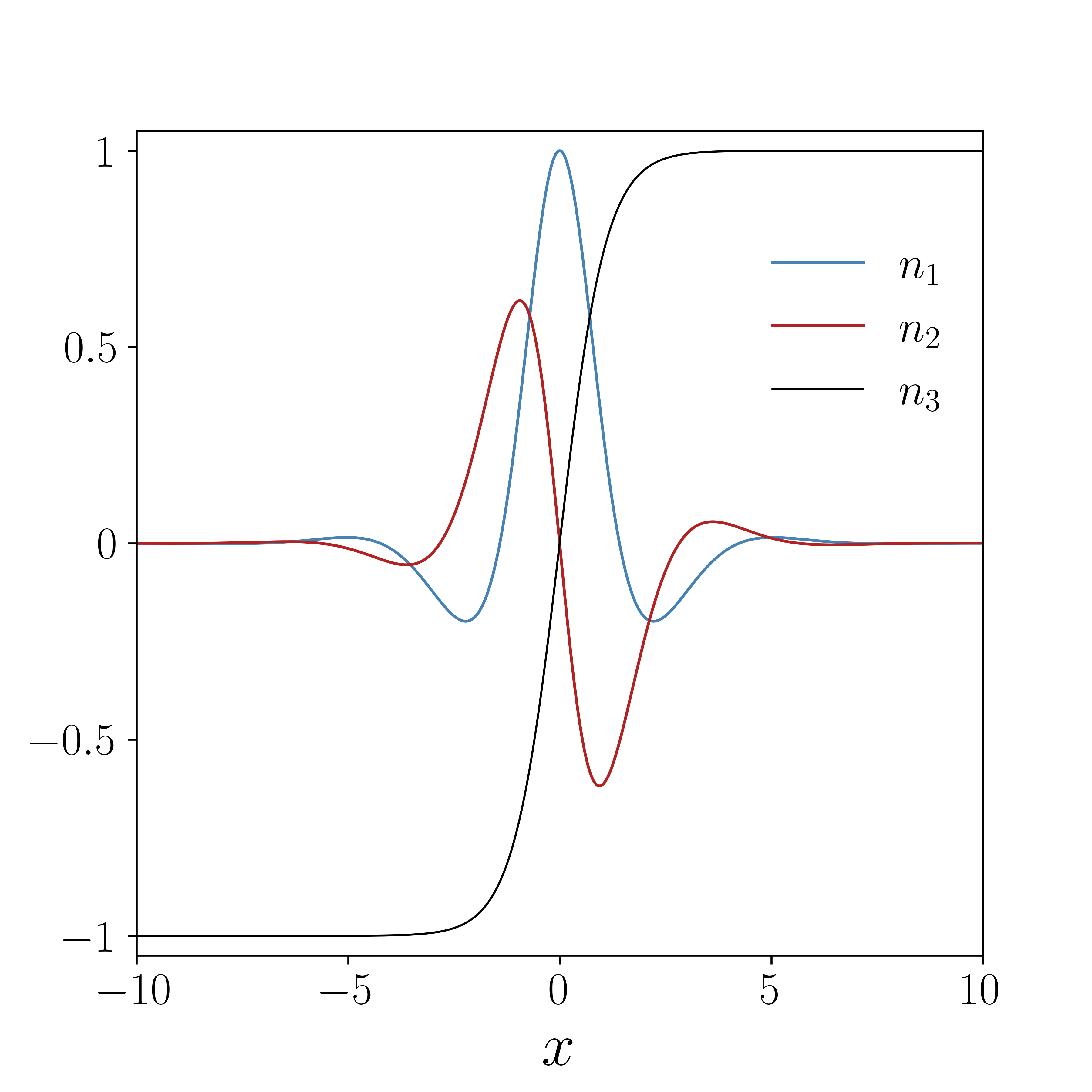}}
\caption{The N\'eel vector components for traveling domain wall solutions for an external field $h=0.5$ and velocities (a) $v=0.6$, (b) $v=0.8$.
We have chosen $x_0=0$ in Eq.~\eqref{eq:ThetaSol} (center of wall at $t=0$) and also $x_1=0$ in Eq.~\eqref{eq:PhiSol}, although these two arbitrary constant need not, in general, be equal.
(We consider here that $\Theta(x=-\infty)=\pi,\; \Theta(x=\infty)=0$.)
}
\label{fig:domainWalls}
\end{figure}

Multiply Eq.~\eqref{eq:ThetaPhi_traveling_Theta} by $\Theta'$ and \eqref{eq:ThetaPhi_traveling_Phi} by $\sin\Theta\,\Phi'$ and then add the results to obtain
\begin{equation} \label{eq:Theta_integrated}
\left[ \Theta'^2 - \left( \frac{1-\hext^2}{1-v^2} - \Phi'^2 \right) \sin^2\Theta \right]' = 0 \Rightarrow
\Theta'^2 - \left( \frac{1-\hext^2}{1-v^2} - \Phi'^2 \right) \sin^2\Theta = 0.
\end{equation}
Substituting Eq.~\eqref{eq:PhiSol}, we have
\begin{equation}
\Theta'^2 - \frac{1-\hext^2-v^2}{(1-v^2)^2} \sin^2\Theta = 0
\end{equation}
and this has the traveling domain wall solution
\begin{equation} \label{eq:ThetaSol}
\tan\left(\frac{\Theta}{2}\right)=e^{\pm (\xi-x_0)/\lw},\qquad \lw = \frac{1-v^2}{\sqrt{1-\hext^2-v^2}}.
\end{equation}
Note that $x_0$, that is, the location of the center of the domain wall at $t=0$, is an arbitrary constant and does not need to be the same as $x_1$ defined in Eq.~\eqref{eq:PhiSol}.
Fig.~\ref{fig:domainWalls} shows the N\'eel vector components for domain walls for two values of the velocity.
Solution \eqref{eq:ThetaSol} requires the assumption
\begin{equation} \label{eq:vc}
\hext^2+v^2 < 1 \Rightarrow |v| < v_c \equiv \sqrt{1-\hext^2}.
\end{equation}
The maximum velocity $v_c$ coincides with the minimum of the phase velocity for spin waves in this system, as can be anticipated by general arguments \cite{2022_AnnPhys_GalkinaKulaginIvanov}.

\begin{figure}[b]
    \begin{center}
    \includegraphics[width=6cm]{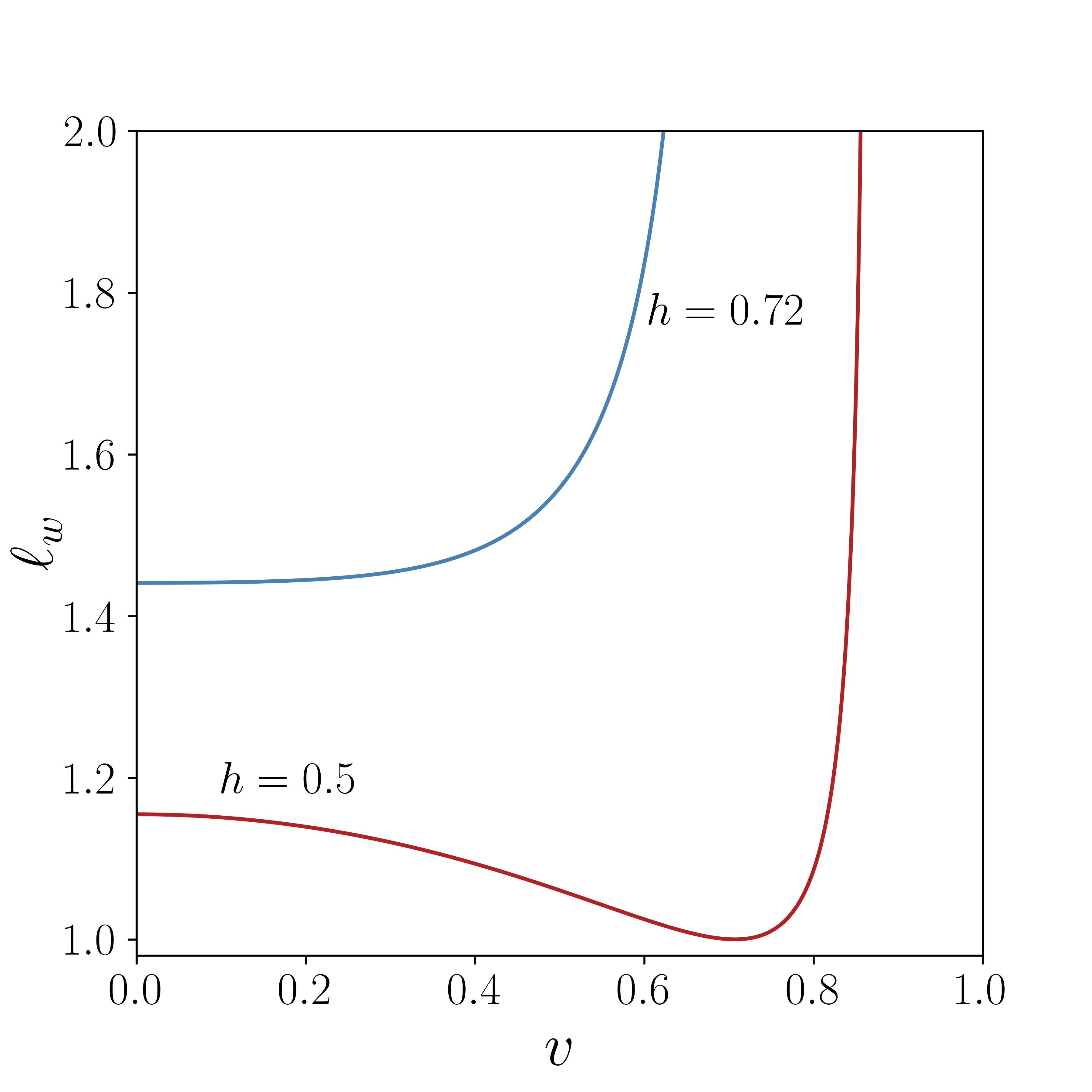}
    \end{center}
    \caption{The width $\lw$ given in Eq.~\eqref{eq:ThetaSol} of traveling domain walls vs velocity for two values of the external field (and $\dm=0$).
    For $\hext < 1/\sqrt{2}$ (e.g., $\hext=0.5$), the width has a minimum at $v=\sqrt{1-2\hext^2}$ and it diverges to infinity at the maximum velocity $v_c=\sqrt{1-\hext^2}$.
    For values of the external field $\hext > 1/\sqrt{2}$ (e.g., $\hext=0.72$), the wall width increases monotonically.}
    \label{fig:wallWidth}
\end{figure}

The functional form of the domain wall for $\Theta$, in Eq.~\eqref{eq:ThetaSol}, is the same as the propagating domain wall solution within the Lorentz-invariant model (for $\hext=0$).
But, while the wall width of the Lorentz invariant model is $\ell=\sqrt{1-v^2}$ and decreases with the velocity $v$ until it vanishes at the maximum velocity $v=1$, the width $\lw$ of the domain wall in the presence of a field behaves radically different.
By a standard expansion close to $v=0$, we can see that $\lw$ initially decreases with velocity for fields $\hext < 1/\sqrt{2} \approx 0.707$.
It has a minimum $\lw=2\hext$ for $v=\sqrt{1-2\hext^2}$ and diverges to infinity at the maximum velocity $v = \sqrt{1-\hext^2}$.
For $\hext > 1/\sqrt{2}$, the wall width is increasing already at $v=0$.
The behavior of the wall width is shown in Fig.~\ref{fig:wallWidth} for a small and for a larger value of the external field.

The linear increase of $\Phi$ given in Eq.~\eqref{eq:PhiSol} means that a full rotation of $\nagn$ in the plane is achieved within a distance $2\pi(1-v^2)/(\hext v)$.
For small velocities, this length scale is large compared to $\lw$ and therefore the effect of N\'eel vector rotation on the plane is not expected to be visible around the central region of the wall.
The effect is larger when $(1-v^2)/\hext v > \lw \Rightarrow v^2 > (1-\hext^2)/(1+\hext^2)$.
This condition allows only for a narrow range of velocities where we can observe domain walls that present full rotation of the in-plane N\'eel component within the domain wall width.

\subsection{Traveling chiral wall}
\label{eq:chiralWalls}

We now include the DM interaction and we will see that propagating domain wall solutions are found in this case, too.
The system of Eqs.~\eqref{eq:ThetaPhi_traveling} is complemented as
\begin{subequations} \label{eq:ThetaPhi_traveling_dm}
\begin{align}
    & (1-v^2)\Theta''-\left[2 \hext v \Phi' + (1-v^2)\Phi'^2+1-h^2\right] \sin\Theta \cos\Theta + 2\dm\sin^2\Theta\,\sin\Phi\,\Phi' = 0 \label{eq:ThetaPhi_traveling_Theta_dm} \\
    & \sin\Theta (1-v^2)\Phi'' + 2 \cos\Theta \Theta' \left[ (1-v^2)\Phi'+\hext v\right] - 2\dm\sin\Theta\,\Theta'\,\sin\Phi= 0. \label{eq:ThetaPhi_traveling_Phi_dm}
\end{align}
\end{subequations}
Equation~\eqref{eq:ThetaPhi_traveling_Phi_dm} can be written as
\begin{equation} \label{eq:Phi_integrated_dm}
 \left[ \sin^2\Theta \left((1-v^2) \Phi' + \hext v \right) \right]' = 2\dm\sin^2\Theta\, \Theta'\,\sin\Phi.
\end{equation}
We set
\begin{equation} \label{eq:phi}
\Phi = k(\xi - x_1) + \phi(\xi)
\end{equation}
and Eq.~\eqref{eq:Phi_integrated_dm} simplifies to
\begin{equation} \label{eq:Phi-small_integrated_dm}
(1-v^2) \left( \sin^2\Theta\,\phi' \right)' = 2\lambda\sin^2\Theta\, \Theta'\,\sin(k(\xi-x_1)+\phi).
\end{equation}
Equation~\eqref{eq:Theta_integrated}, which does not contain a DM term, can still be obtained from Eqs.~\eqref{eq:ThetaPhi_traveling_dm}.
Substituting the form \eqref{eq:phi}, it becomes
\begin{equation} \label{eq:Theta_integrated_phi}
\Theta'^2 - \left[ \frac{1-v^2-h^2}{(1-v^2)^2} - \phi' \left(\frac{2h v}{1-v^2} - \phi' \right) \right] \sin^2\Theta = 0.
\end{equation}

\begin{figure}[t]
\centering
\subfloat[]
{\includegraphics[width=6cm]{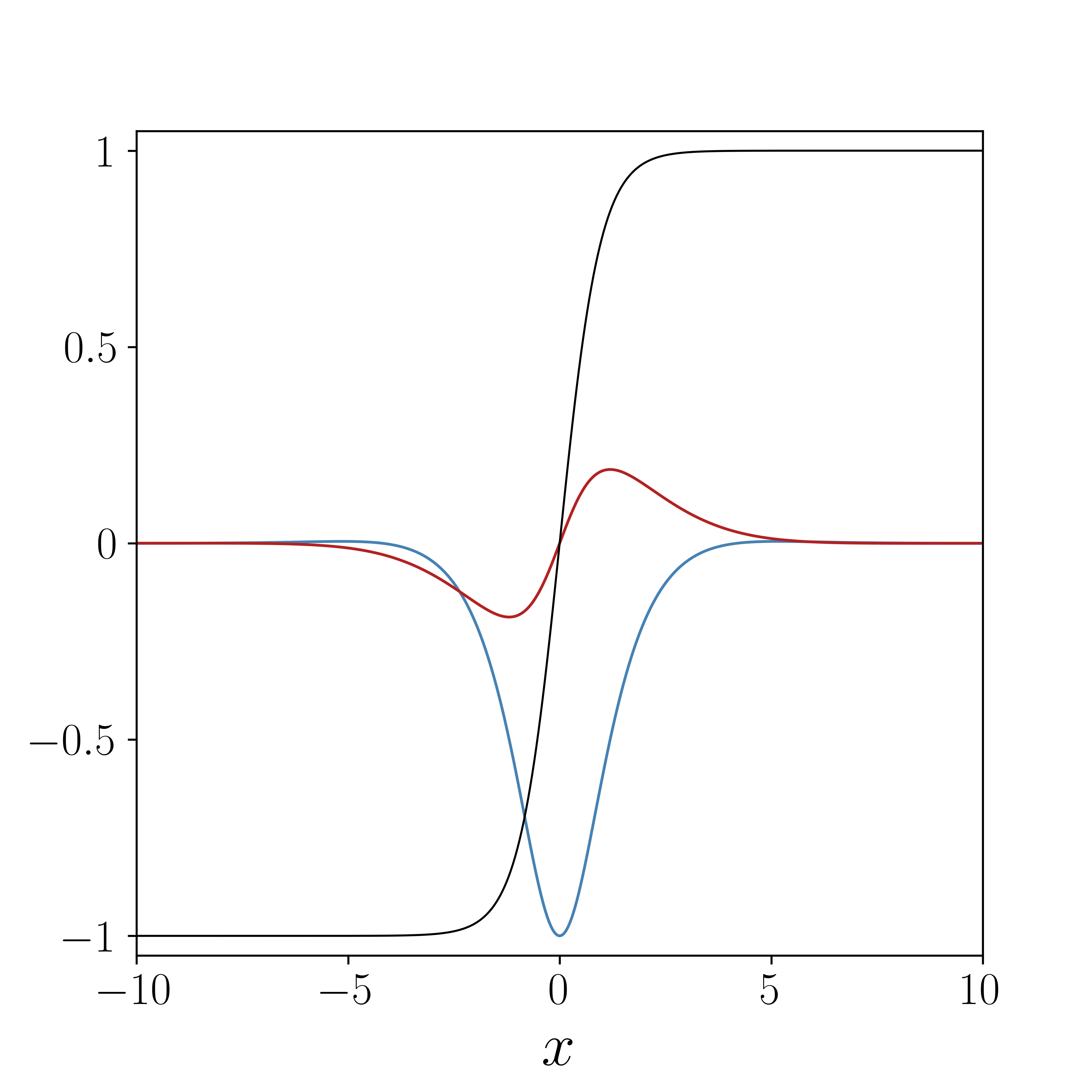}}\qquad
\subfloat[]
{\includegraphics[width=6cm]{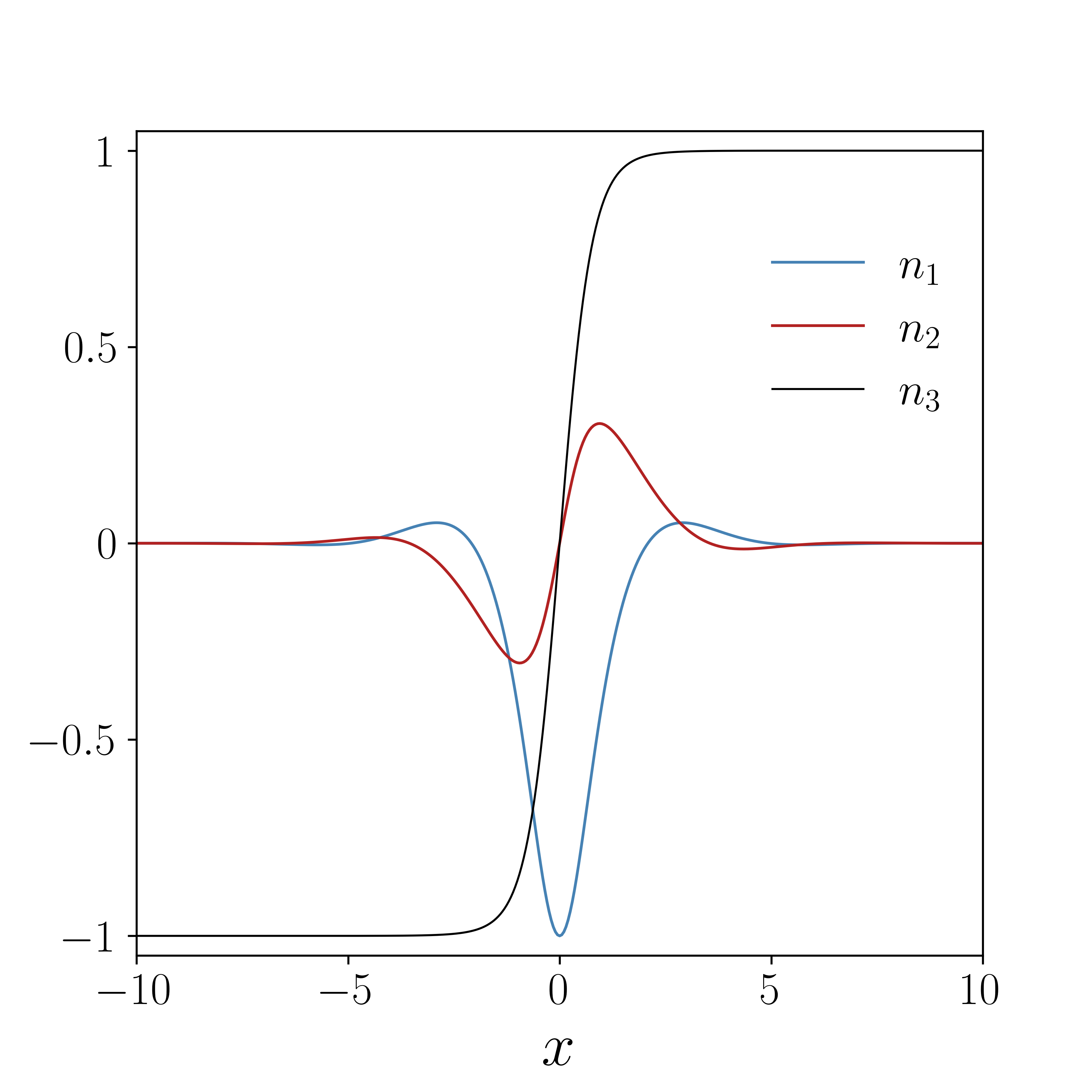}}
\caption{The N\'eel vector components for propagating chiral domain walls for DM parameter $\dm=0.5$, external field $\hext=0.5$, and
velocities (a) $v=0.6$ and (b) $v=0.8$.
The $n_2$ component is smaller in magnitude and the wall width is smaller compared to the non-chiral walls in Fig.~\ref{fig:domainWalls}.
For the chiral wall, we necessarily have $x_0=x_1$, i.e., the center of the wall ($n_3=0$) is at $\Phi=0$.
(We consider here that $\Theta(\xi=-\infty)=-\pi,\; \Theta(\xi=\infty)=0$.)
}
\label{fig:domainWalls_dm}
\end{figure}

To proceed, we write, for small $\dm$,
\begin{equation}
    \phi = \lambda \phi_1 + O(\lambda^2),\qquad
    \Theta = \Theta_0 + O(\dm).
\end{equation}
Then Eq.~\eqref{eq:Theta_integrated_phi} implies that $\Theta_0$ is the domain wall \eqref{eq:ThetaSol}, and Eq.~\eqref{eq:Phi-small_integrated_dm} gives, to order $O(\lambda)$,
\begin{equation} \label{eq:phi-small_integral}
 \sin^2\Theta_0\, \phi_1' = \frac{2}{1-v^2} \int \sin^2\Theta_0 \Theta_0'\,\sin(k(\xi-x_1))\, d\xi.
\end{equation}
Far from the wall center, it is 
\begin{equation}
    \sin\Theta_0 \approx \frac{1}{2}e^{-|\xi|/\lw},\qquad
\Theta_0' =\frac{1}{\lw} \sin \Theta_0 \approx \frac{1}{2|\lw|}  e^{-|\xi|/\lw},\qquad |\xi| \gg \lw.
\end{equation}
Using these forms in Eq.~\eqref{eq:Phi-small_integrated_dm}, it implies that $\phi' \to 0$ at spatial infinity.
Going now to Eq.~\eqref{eq:phi-small_integral}, we conclude $x_1=x_0$ or $x_1=x_0\pm\pi/k$, so that the integrand is odd around the center $x_0$ of the domain wall and therefore $\phi_1'$ can go to zero at both $\xi\to\pm\infty$.

Fig.~\ref{fig:domainWalls_dm} shows the domain wall solutions for parameter value $\dm=0.5$ and an external field $\hext=0.5$.
We have verified numerically that any initial condition eventually converges to give $\Phi=0$ in the center of the domain wall, which corresponds to $x_0=x_1$ (if we assume $\Theta(\xi=0)=-\pi/2$).
The DM interaction acts to quench the variation of $\Phi$ in the central part of the wall, keeping its value closer to $\Phi=0$ (compared to the non-DM case), so it favors a definite chirality of the wall.
This means that the $n_2$ component is suppressed in the chiral case shown in Fig.~\ref{fig:domainWalls_dm} compared to the corresponding walls in the non-chiral case shown in Fig.~\ref{fig:domainWalls}.
This confirms the result in Eq.~\eqref{eq:phi-small_integral}.
Also, the wall width is seen in Fig.~\ref{fig:domainWalls_dm} to be smaller compared to the non-chiral case.

\begin{figure}[t]
    \begin{center}
    \includegraphics[width=6cm]{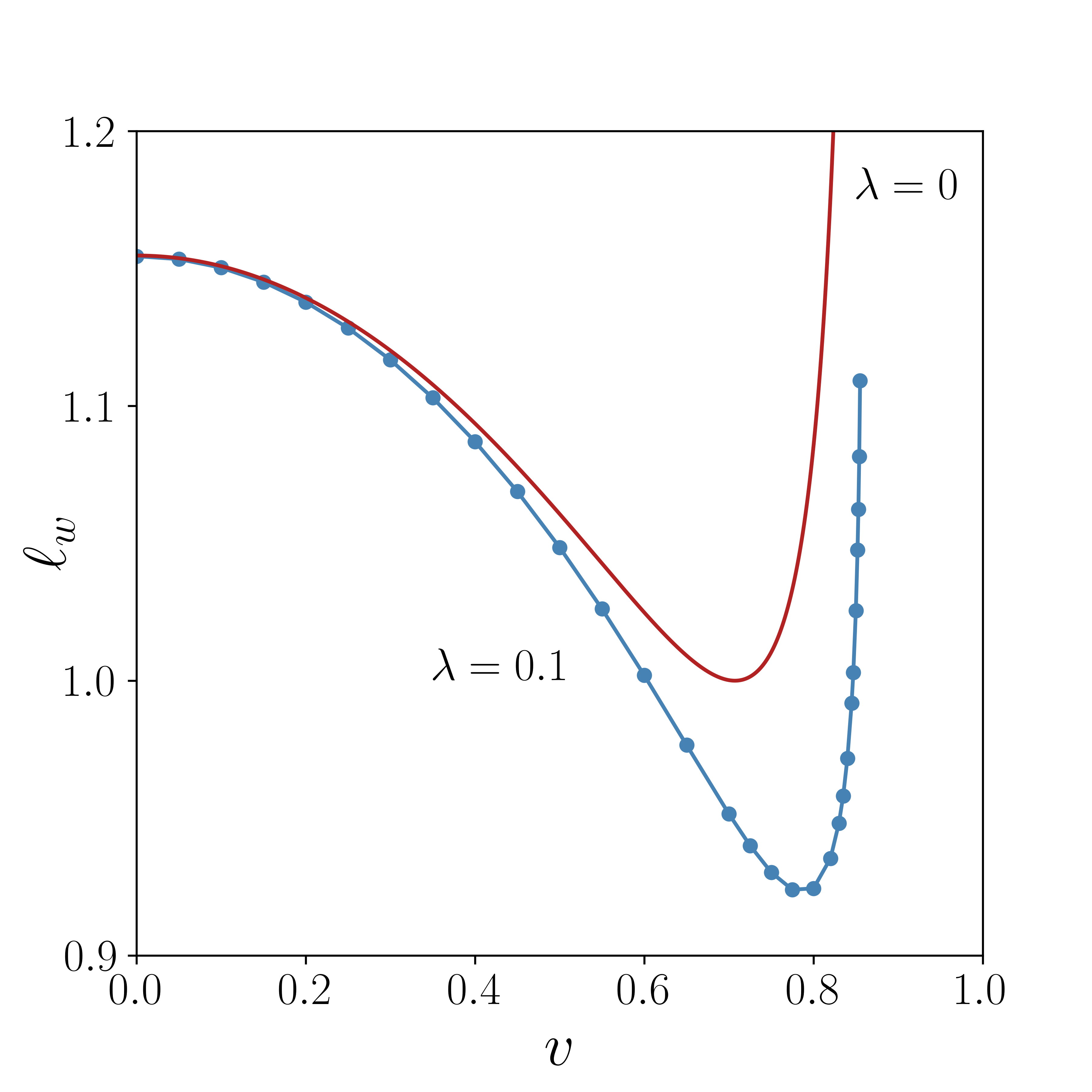}
    \end{center}    
    \caption{The width $\lw$ for traveling domain walls vs velocity for $\lambda=0.1$ and field $\hext=0.5$ is shown by dots connected with a line.
    The dots for $\lw$ have been found by fitting a hyperbolic tangent to the numerical solutions.
    The corresponding curve for $\dm=0$ is plotted for comparison (it is identical to the curve in Fig.~\ref{fig:chiralwallWidth}).
}
    \label{fig:chiralwallWidth}
\end{figure}

Fig.~\ref{fig:chiralwallWidth} shows the width of the chiral DW as a function of the velocity for $\lambda=0.1$, while the wall width for $\lambda=0$ is also plotted for comparison.
In both cases, we have an external field $\hext=0.5$.
For the case $\dm\neq 0$, we calculated the thickness of DW by fitting a hyperbolic tangent to the numerical solutions.
The qualitative behavior is similar in the two cases, but the minimum is pronounced when $\dm\neq 0$.
The results in Fig.~\ref{fig:chiralwallWidth}, confirm that the wall width in the chiral case is smaller compared to the non-chiral case.

\section{Propagating spiral}
\label{sec:spiral}

One may ask about the fate of the domain wall when this is forced to propagate at a velocity higher than the maximum velocity, $v > v_c$, given in Eq.~\eqref{eq:vc}.
In this section, we confine our study to $\dm=0$.
We start this investigation by writing a functional $F$, such that its minimization gives Eq.~\eqref{eq:sigmaModel_1D_traveling}.
It is
\begin{equation} \label{eq:energy_traveling}
\Eeff = \frac{1-v^2}{2} \int \nagn'^2\,dx + \hext v \int \e_3\cdot (\nagn\times\nagn')\,dx + \frac{1-\hext^2}{2} \int (1-n_3^2)\,dx.
\end{equation}
The minimizing $\nagn$ is a solution of the original equation under the constraint of constant velocity.
This indicates that $F$ should be equivalent to the energy functional augmented with a Lagrange multiplier $v$ times the linear momentum.

Assume now an in-plane configuration
\begin{equation} \label{eq:inplane_periodic}
n_1 = \cos\Phi,\quad n_2 = \sin\Phi,\quad n_3 = 0
\end{equation}
which reduces the functional to
\[
\Eeff =  \int \left(\frac{1-v^2}{2} \Phi'^2 + \hext v \Phi' + \frac{1-\hext^2}{2} \right) dx.
\]
The minimum of $F$ is obtained for
\begin{equation} \label{eq:spiral_Phi_prime}
\Phi' = -\frac{\hext v}{1-v^2}
\end{equation}
giving the integrand in Eq.~\eqref{eq:energy_traveling},
\[
f = \frac{1}{2} \frac{1-\hext^2-v^2}{1-v^2}.
\]
This shows that the spiral state \eqref{eq:inplane_periodic} achieves a lower minimum, $F < 0$, than the polarized state $\nagn=\pm \e_3$ when $\hext^2+v^2 > 1$, i.e., for $v > v_c$.
We have verified numerically that the damped equation of motion under the constraint of a velocity $v>v_c$ leads to the spiral.
One could describe the physical picture that emerges for $v>v_c$ in two alternative ways. 
One way is to consider that the polarized domains, on either side of the domain wall, are destabilized to give the spiral \eqref{eq:inplane_periodic}, and another way is to consider that the central region of the domain wall extends to occupy the whole space as $v\to v_c$ and $\lw\to\infty$. 
The spiral \eqref{eq:spiral_Phi_prime} gives a minimum of $F$ in the velocity range $\sqrt{1-\hext^2} < |v| < 1$, and its period is $2\pi (1-v^2)/(\hext v)$.
At the critical velocity $v=v_c$, the period is $2\pi\sqrt{1-\hext^2}/\hext$, while at $v\to 1$ the period goes to zero, giving the singular behavior typically expected for Lorentz invariant propagating solutions.

The spiral solution in Eq.~\eqref{eq:inplane_periodic} is mainly due to the presence of the second term in the functional \eqref{eq:energy_traveling} which originates in the dynamical term due to the external field.
This term has the same form as a chiral term with DM vector along $\e_3$.
The spiral solution could thus have been anticipated on the basis of the same arguments as for the well-known spiral of chiral magnetic models \cite{1994_JMMM_BogdanovHubert,2002_PRB_ChovanPapanicolaou}.

\begin{figure}[t]
    \begin{center}
    \includegraphics[width=6cm]{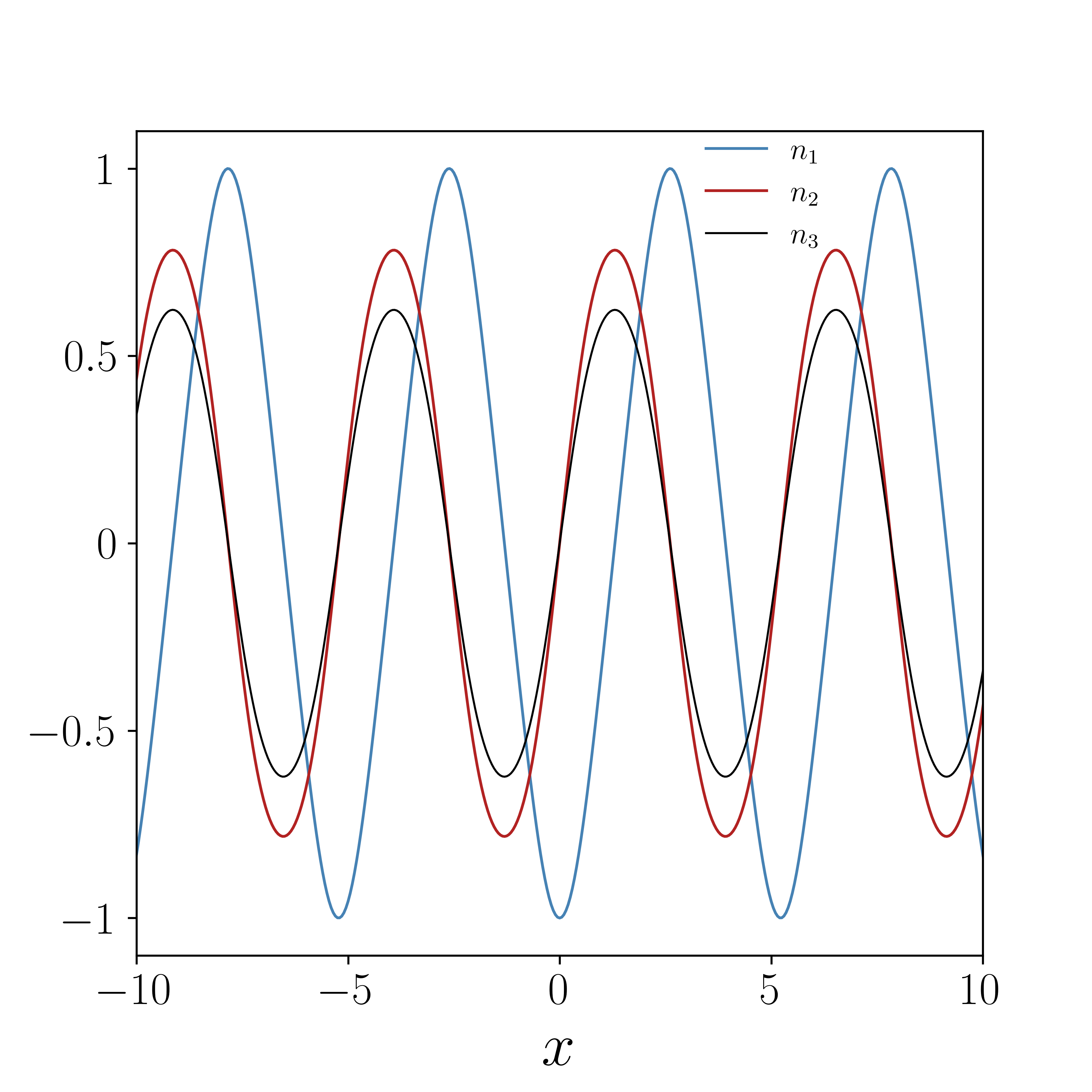}
    \end{center}    
    \caption{A spiral propagating with velocity $v=0.79$ for parameters $\hext=0.6$ and $\dm=0.2$.
    This velocity is below $v_c=0.8$ which is the critical velocity for the transition to the spiral when $\dm=0$.
}
    \label{fig:spiralDM}
\end{figure}

If we add a DM term in the model, we find numerically that the spiral is significantly modified.
All three components of $\nagn$ oscillate, as shown in Fig.~\ref{fig:spiralDM}.
We have also found that the spiral is obtained for velocities somewhat lower than the value $v_c$.
For example, for $\hext=0.6$, we have $v_c=0.8$.
For $v=0.79$, we obtain a domain wall for $\dm=0$ but this becomes a spiral when we increase $\dm$ (this case is shown in Fig.~\ref{fig:spiralDM}).
This means that the DM reduces the maximum velocity of the domain wall.

The results of this section indicate that, if a method (experimental or simulation) were developed to force the magnetic configuration, under model \eqref{eq:sigmaModel_1D_traveling}, to propagate with high velocity, then the spiral would appear spontaneously and its period would decrease with increasing velocity.

\section{Concluding remarks}
\label{sec:conclusions}

We have studied solitary wave motion of domain walls in antiferromagnets under a magnetic field parallel to the easy axis.
The magnetic field introduces a dynamical term in the model that breaks the Lorentz invariance and generates some special features of the propagating domain walls.
The maximum possible velocity is reduced due to the magnetic field.
For stronger fields, the domain wall width is a non-monotonic function of the velocity and it presents a minimum.
The wall width diverges to infinity at the maximum velocity for all fields.
When a DM interaction term is added to the model, the domain wall profile is modified, and it acquires a chiral character although it does not become a fully chiral wall.
If the system is driven to obtain a velocity higher than the maximum velocity allowed for a domain wall, it is expected that a propagating spiral will be spontaneously created with the N\'eel vector lying fully in-plane perpendicular to the easy-axis.
When there is a DM interaction, the spiral is more complicated with all three magnetization components being nonzero and periodic in space.

\acknowledgments

The authors gratefully acknowledge discussions with B.A. Ivanov which helped improve the quality and completeness of the present work.

\bigskip
\bibliographystyle{apsrev4-1.bst}
\bibliography{references.bib}

\end{document}